# Advancement of LANSCE Accelerator Facility as a 1-MW Fusion Prototypic Neutron Source

Yuri K. Batygin, Eric J. Pitcher, LANL, Los Alamos, NM 87545, USA

Abstract

The Fusion Prototypic Neutron Source (FPNS) is considered to be a testbed for scientific understanding of material degradation in future nuclear fusion reactors [1-3]. The primary mission of FPNS is to provide a damage rate in iron samples of 8-11 dpa/calendar year with He/dpa ratio of ~10 appm in irradiation volume of 50 cm$^3$ or larger with irradiation temperature 300 – 1000º C and flux gradient less than 20%/cm in the plane of the sample. The Los Alamos Neutron Science Center (LANSCE) is an attractive candidate for the FPNS project. The Accelerator Facility was designed and operated for an extended period as a 0.8-MW Meson Factory. The existing setup of the LANSCE accelerator complex can nearly fulfill requirements of the fusion neutron source station. The primary function of the upgraded accelerator systems is the safe and reliable delivery of a 1.25-mA continuous proton beam current at 800-MeV beam energy from the switchyard to the target assembly to create 1 MW power of proton beam interacting with a solid tungsten target. The present study describes existing accelerator setup and further development required to meet the needs of FPNS project.

## 1. Introduction

The Fusion Prototypic Neutron Source is a proposed facility under consideration by the fusion materials community to irradiate materials under conditions similar to that in fusion reactor. It is a necessary step in establishing an adequate knowledge of material degradation in a D-T fusion environment. The FPNS will provide required materials data to build more advanced fusion facilities, such as the Fusion Nuclear Science Facility [4], DEMO [5], and the Chinese Fusion Engineering Test Reactor [6].

One option under study to meet the demanding FPNS radiation damage rate in iron of at least 8 dpa/year is a spallation neutron source, which could be constructed at the Los Alamos Neutron Science Center. Studies proposing the use of LANSCE for fusion materials irradiations date back to 1981 [7, 8]. The LANSCE accelerator started routine operation in 1972 as a 0.8 MW average proton beam power facility for meson physics research, and delivered high-power beam for a quarter century [9]. The accelerator currently delivers beams to five experimental areas (see Figure 1 and Table 1). The accelerator is equipped with two independent injectors for H$^+$ and H$^-$ beams, merging at the entrance of a 201.25 MHz Drift Tube Linac (DTL). The DTL performs acceleration up to the energy of 100 MeV. After the DTL, the Transition Region beamline directs a 100 MeV proton beam to the Isotope Production Facility, while the H$^-$ beam is accelerated up to the final energy of 800 MeV in an 805-MHz Coupled Cavity Linac. The H$^-$ beams, created with different time structure by a low-energy chopper, are distributed in the Switch Yard (SY) to four experimental areas: the Lujan Neutron Scattering Center equipped with a Proton Storage Ring

(PSR), the Weapons Neutron Research facility (WNR), the Proton Radiography facility (pRad), and the Ultra-Cold Neutron facility (UCN).

The design proposed to upgrade the LANSCE facility for FPNS project employs an innovative annular target with the fusion materials irradiation region occupying the central flux trap inside the annular target [10]. Estimations of interaction of 1 MW proton beam with the target indicate the ability to generate a neutron flux of $10^{15}$ n cm$^{-2}$ s$^{-1}$, which is sufficient to generate a damage rate in iron of 20.6 dpa per full-power year averaged over a 53-cm$^3$ volume. The calculated He-to-dpa ratio in this volume is 14.6 appm/dpa, near the desired value of 10 appm/dpa. The cost-effective solution of the proposed FPNS at LANSCE is based on the availability of accelerator and "Area A" experimental hall infrastructure with minimal required upgrade needs.

## 2. Accelerator Setup

The basic parameters of FPNS beam setup are presented in Table 2. The accelerator should safely and reliably deliver an average 1.25-mA proton beam current at 800-MeV beam energy from the switchyard to the FPNS target located in Area A. The rastering system should have the capability to raster the proton beam within a thin ring area at the target of volume of 50 cm$^3$ with time-averaged beam current density non-uniformity of less than ±3%. Spatial dimensions of the rastered beamlet on target (including beamlet spreading resulting from beam jitter in the linac) should be minimized, with capability to image the beam spot on target. A low-power (20 kW) tuning beam dump is required at the target area to provide the capability to quickly tune the beam for delivery to FPNS. In order to provide safety margins in the rastering system, the design presented in this paper assumes rastering area with minimal radius at the target $R_{min}$ = 5.5 cm and maximum radius of $R_{max}$ = 6.9 cm.

A unique feature of the LANSCE accelerator facility is simultaneous acceleration of multiple beams. In addition to the five existing experimental areas, the FPNS target would be placed in the existing Area A experimental hall, which served previously as a meson physics research area. For FPNS operation, the time structure of the LANSCE accelerator needs to be rearranged (see Figure 2 and Table 3). The Lujan Center would continue receiving 20 Hz x $625\,\mu s$ pulses of H$^-$ beam with 10 mA/bunch peak current, which is translated into beam with average current of 100 $\mu A$ and average beam power of 80 kW. The further upgrade of this beam up to $125\,\mu A$ and beam power of 100 kW would be achievable with an increase of the beam pulse width from $625\,\mu s$ to $781\,\mu s$, or/and with an upgrade of the H$^-$ ion source delivering higher peak current. FPNS beam is operated at 78 Hz x $850\,\mu s$ pulse pattern with 18.87 mA per bunch H$^+$ peak current producing a beam with average current of 1.25 mA and average beam power of 1 MW. Operation with larger pulsed beam current of 21 mA was successfully demonstrated as a test for the proposed LANL Long Pulse Spallation Source, delivering H$^+$ beam with energy 800 MeV, average current 315 µA and reliability of ~ 88% for 12 hours [11]. FPNS beam shares the same RF pulses (that is, both beams are accelerated simultaneously) with Weapons Neutron Research facility (WNR) beam, which is operated at 100 Hz x $850\,\mu s$ pattern delivering H$^-$ beam with average current of 5.95 $\mu A$ and average power of 4.76 kW to the WNR target. H$^-$ beam delivered to the Ultra Cold Neutron (UCN) research facility consumes 8 Hz x $850\,\mu s$ every 5 sec, which is approximately equivalent to 2 Hz of continuous operation. UCN beam "steals" these cycles from FPNS beam, and is accelerated in the same RF pulses with WNR beam. H$^-$ beam delivered to Proton Radiography

Facility "steals" single $850\,\mu s$ macro-pulses of H⁻ beam from WNR beam. The remaining 20 Hz of RF pulses are consumed by H⁺ beam delivered to the Isotope Production Facility (IPF). That beam with the same peak current of 18.87 mA as FPNS beam is accelerated at 20 Hz x $662\,\mu s$ pulse pattern up to an energy of 100 MeV in the DTL only. Currently IPF beam is supposed to deliver H⁺ beam with average current of $250\,\mu A$ and energy of 100 MeV or an average power of 25 kW to the IPF target. A potential upgrade of the IPF facility to an average beam power up to 31 kW is achievable with an increase of IPF beam pulse width to $826\,\mu s$.

The performance of the RF system is vital for accelerator upgrade as a 1 MW average beam power facility. Recent completion of the LANSCE Risk Mitigation Project [12] included significant upgrade of RF power feeding the DTL. Three out of four 201.25 MHz amplifiers (Tanks 2 - 4) were replaced with newly developed RF power systems based on TH628L Diacrodes [13] (see Figure 3). Replacement of the amplifier of Tank 1 for a more powerful one is scheduled for winter 2020. In addition, a new digital low-level RF (dLLRF) control system was installed, and end-of-life coupled-cavity linear accelerator (CCL) klystrons were replaced to insure further stable beam operation (see Figure 4). Upgrade of dLLRF will be required to accommodate both existing feedforward and adaptive feedforward modes working through the CCL. Also, dLLRF for bunchers should be installed.

Operation beam duty factor is a combination of that of the most powerful beams delivered to the various experimental areas:

$$\text{(H}^+\text{ FPNS) 80 Hz x 850 μs +(H}^+\text{ IPF) 20 Hz x 662 μs +(H}^-\text{ Lujan) 20 Hz x 625 μs} = 9.4\%. \quad (1)$$

The Drift Tube Linac RF gate requires an additional 250 μs to each beam gate for Low Level RF control system to be locked and stabilize amplitudes and phases of DTL tanks within the level of 1%, 1° correspondingly. It results in the following value for DTL RF Duty Factor:

$$\text{(H}^+\text{ FPNS) 80 Hz x 1100 μs +(H}^+\text{ IPF) 20 Hz x 912 μs +(H}^-\text{ Lujan) 20 Hz x 875 μs} = 12.4\%. \quad (2)$$

The Coupled-Cavity Linac requires 120 Hz x 950 μs RF pulse setup for all beams, which translates to a CCL Duty Factor of 11.4%. The DTL Duty Factor limit is 15% [14] and that of the CCL is 12.0% [15], therefore the proposed setup is within the acceptable limits. Thermal and structural analysis of Drift Tube Linac was done in Refs. [16, 17], and that of Coupled-Cavity Linac in Ref. [18].

Acceleration of the beam with higher peak current results in additional power consumption by the beam. Beam loading is determined by the power $P_b$ consumed by the beam with current $I$ in the accelerating section receiving energy gain $\Delta W$:

$$P_b = I\,\Delta W. \quad (3)$$

RF power $P_{RF}$ required from power amplifiers or klystrons to feed the accelerating section with voltage $U$, transit time factor $T$, and effective shunt impedance $R_{sh}$ is determined by

$$P_{RF} = \frac{(UT)^2}{R_{sh}} + P_{loss}, \quad (4)$$

where $P_{loss}$ is the power loss in connecting waveguides and RF bridges. Total power, $P_{tot} = P_{RF} + P_b$, should remain smaller than limited power, $P_{lim}$. The largest value of beam peak current out of 6 experimental areas is the FPNS proton beam current of 18.87 mA. In the Drift Tube Linac, energy gains per tank are 4.6447 MeV (Tank1), 35.9406 MeV (Tank 2), 31.3969 MeV (Tank 3) and 27.283 MeV (Tank 4). In the CCL, energy gain per module is between 14 – 16 MeV. Tables 4, 5 contain RF parameters of the DTL and CCL with FPNS proposed beam setup.

Another important topic in consideration of a high-power upgrade is associated with beam loss. Beam losses at LANSCE are controlled by various types of loss monitors. The main control is provided by Activation Protection (AP) detectors, which are one-pint sized cans with a photomultiplier tube immersed in scintillator fluid. AP detectors integrate the signals and shut off the beam if the beam losses around an AP device exceed 100 nA of average current. The same devices are used as beam loss monitors (LM), where the signal is not integrated and therefore one can see a real-time of beam loss across the beam pulse.

Another type of loss monitor is the Ion Chamber (IR) detectors. These are used in the high energy transport lines. They are usually located in concert with Gamma Detectors (GDs) that feed into the Radiation Safety System. An advantage of the IRs is that they do not saturate at high loss rates like the AP devices.

The third type of beam loss monitor is the Hardware Transmission Monitor (HWTM). The HWTM system measures the beam current losses between current monitors and terminates operation if a difference between current monitors exceeds a threshold value. The HWTM system and the activation protection system incorporated with the fast protect system provide a means to minimize activation to areas that require "hands on" maintenance. The current monitor toroids are usually a one hundred or a two hundred turns toroid influenced by the charged particle beam. A single turn loop is included to send a calibration pulse through to make sure the current monitor is operational. The transmission data differences derived from the toroid signals provides information on the beam losses. The typical HWTM constraints of 10 nA/m of accelerator length are aimed to prevent moderate to large beam losses usually occur during accelerator tune-up period or when beamline device malfunctions.

Long-term operation of 800-MeV, 80-kW $H^-$ beam indicates that summed losses of that beam in the high-energy part of the linac do not exceed 200 nA. Losses of $H^-$ beam are mostly determined by stripping of $H^-$ ions on residual gas, intra-beam stripping, and Lorentz stripping. Beam losses of $H^+$ beam are typically one order of magnitude smaller than that of $H^-$ beam at the same current due to the absence of stripping mechanisms [19]. Therefore, we can expect that losses of 1-MW $H^+$ beam will not exceed losses of 80-kW $H^-$ beam. Previous operation of the accelerator facility at 0.8 MW indicated that losses of $H^+$ beam were noticeably smaller than that of lower-current $H^-$ beam.

## 3. High-Energy Beam Transport

The beamline from the LANSCE switchyard to the FPNS target is composed of existing Switchyard - Line A beamline and additional beamline, which is supposed to be constructed in Area A (see Figures 5 - 7). The FPNS target will be located near the center of Area A, some 100 m downstream of the end of the 800-MeV accelerator. Figure 5 shows a sketch of the beamline that will deliver 800-MeV protons from the accelerator to the FPNS target front face.

Presently, Line A transports H⁻ particles bound for PSR and WNR through quadruple doublet LAQM01, LAQM02 into Line D. In the past, protons were transported to Area A, and the H⁻ particles and protons were separated in the bending magnet LABM01. This will again be the case when FPNS comes on line. Thus, the settings of the first two quadrupoles of Line A are determined by Line D requirements and cannot be altered. The present Line A is preserved in magnetic-element placement through quadrupole doublet LAQM05-LAQM06. The new beamline in Area A has the minimum number of quadrupoles (LAQM07 - LAQM08) necessary to meet all beam-optics requirements (beam parameters at the target, and beam-centroid excursions at the target). The raster-magnet section has two raster magnets LAHR01, LAVR01. An additional beam-optics requirement is driven by the need to shield the upstream beamline from backstreaming neutrons. It is achieved by placing a shielding wall with a small aperture upstream of the target. There also is a beamline spur, composed of two 15° dipoles (LABM05 and LABM06), leading to a tune-up beam dump.

The beam-at-target requirements are dictated by neutronics considerations. A round beam with minimal spot is painted onto an annular target. Beam-centroid jitter is expected to be at the level of 0.5 rms of beam size. The raster scheme is sufficiently flexible to accommodate changes in the size of the painted annular spot as the design of the FPNS target evolves. For example, a ring with minimum radius of $R_{min}$ = 5.5 cm and maximum radius of $R_{max}$ = 6.9 cm has been assumed in developing the beam optics design presented here, but this design can easily be adjusted to paint a ring with $R_{min}$ = 3.2 cm and $R_{max}$ = 4.3 cm, as assumed in the neutronics design [10]. To be conservative, the beamline apertures are dimensioned to accommodate beam-centroid excursions at the target of up to 10 cm, both horizontally and vertically.

Application requires minimization of the rastered beam spot on target, while keeping 7-rms beam size within existing beampipe radius of 7.62 cm (or one-rms beam radius within 1 cm). The target area will be surrounded with shielding to prevent irradiation of upstream equipment by backscattered neutrons. The distance from the upstream shield wall to the target is selected to be 10 m. An additional 4 m of beamline is required for placement of raster magnets. Therefore, the total distance from the exit of the last quadrupole to the target is 14 m. Evolution of beam size in the drift space, $R$, is determined by the beam envelope equation:

$$\frac{R}{R_t} = \sqrt{(1+\frac{R'_t}{R_t}z)^2 + (\frac{\ni}{R_t^2})^2 z^2} \,, \tag{5}$$

where $R_t, R'_t$ are beam envelope size and slope at the target, $\ni$ is the unnormalized beam emittance, and $z$ is the drift space. Equation (5) determines enhancement of beam size from the target to the upstream beamline. To minimize beam size at the drift, the slope of beam envelope at the target should be

$$R'_t = -\frac{R_t}{z}. \tag{6}$$

Then, the beam size at the target, $R_t$, is related to beam size $R$ at the distance $z$ upstream of the target as

$$R_t = \frac{Э}{R} z . \tag{7}$$

Keeping in mind that maximal rms beam size in the beamline is limited to 1 cm, we select a value of $R = 0.7$ cm in order to accommodate additional variation of beam envelopes in the last quadrupole doublet before the target within the 1-cm limit. For the beam with typical unnormalized rms beam emittance after the linac $Э = 0.05 \, \pi$ cm mrad, the rms beam size and slope at the target is $R_t = 1$ mm, $R'_t = -7.14 \cdot 10^{-5}$.

Figure 8 illustrates results of a 6D beam dynamics simulations of FPNS beam using the BEAMPATH code [20]. The quadrupole doublet LAQD01-02 is common for both H$^+$ and H$^-$ beams. The values of quadrupole gradients in that doublet of 1.5 kGs/cm are selected from a regular run of H$^-$ beam. Four bending magnets LABM01-04 separate vertically the H$^+$ and H$^-$ beams after linac. They were used at the time of simultaneous operation of both beams at the energy of 800 MeV. Quadrupole doublets LAQD03-04 and LAQD05-06 were part of the 0.8-MW H$^+$ beam transport to Area A. Those magnets, together with existing beam diagnostics, proved to be effective for providing reliable high-energy beam transport. An additional quadruple doublet LAQD07- LAQD08 described above provides the desired small beam size at the target. The values of quadrupole gradients were optimized using code TRACE-2D [21]. Table 6 lists the beamline elements, in beam-line order, from the downstream end of the vertical bend to the target front face.

Simulations were performed starting from the exit of the linac with realistic beam parameters. After the linac, beam is extracted with unnormalized emittance of $Э = 0.05 \, \pi$ cm mrad and Twiss parameters $\alpha_{x,y} \approx -0.2...-0.4$; $\beta_{x,y} \approx 1...2$ cm/mrad. Typical measured FWHM beam energy spread after the linac is $dW/W = 1.8 \cdot 10^{-3}$. Longitudinal bunch size is measured at the energy 211 MeV to be 25° at 805 MHz. During acceleration the bunched beam size drops due to the phase damping mechanism as

$$\Delta \varphi \sim \frac{1}{(\beta \gamma)^{3/4}} , \tag{8}$$

and becomes 10° at the energy of 800 MeV. Figures 9, 10 contain beam distributions in phase space after the linac and at the target.

## 4. Beam Rastering

The purpose of beam rastering is to uniformly irradiate the target within a ring area of $R_{min} < R < R_{max}$. Rastering is achieved through sinusoidal variation of the deflecting field in a set of two raster (steering) magnets in horizontal and vertical directions. Fields in the magnets are shifted by 90° to perform circular irradiation, therefore the 2D beam centroid variation is determined as

$$\begin{pmatrix} R_x \\ R_y \end{pmatrix} = R(t) \begin{pmatrix} \cos(\omega t + \varphi_o) \\ \sin(\omega t + \varphi_o) \end{pmatrix} , \tag{9}$$

where $R(t)$ is the beam radius at the target, and $\omega = 2\pi f$ is the circular frequency of raster. In order to achieve uniformity of the irradiated region, an increment of area covered by the beam after each turn, $dS$, is kept constant [22]:

$$dS = 2\pi R(t) dR(t) = const. \tag{10}$$

To determine the dependence of beam radius at the target, let us divide the left and right parts of Eq. (10) by $d\varphi = \omega dt$, where $\varphi$ is the azimuth angle:

$$\frac{2\pi}{\omega} R(t) \frac{dR(t)}{dt} = \frac{dS}{d\varphi}. \tag{11}$$

Integration of Eq. (11) provides the dependence of radius on time of irradiation

$$R^2(t) - R_{min}^2 = \frac{dS}{d\varphi} \frac{\omega}{\pi} t. \tag{12}$$

The unknown value of $(dS/d\varphi)(\omega/\pi)$ is determined from the condition that after time of irradiation $\tau$, the beam radius reaches the maximum value of $R_{max}$:

$$\frac{dS}{d\varphi} \frac{\omega}{\pi} = \frac{R_{max}^2 - R_{min}^2}{\tau}. \tag{13}$$

Combining Eqs. (12) and (13), the dependence of beam radius on time of irradiation is given by

$$R(t) = R_{min} \sqrt{1 + (\frac{R_{max}^2}{R_{min}^2} - 1) \frac{t}{\tau}}. \tag{14}$$

Uniform irradiation is achieved by overlapping multiple beam turns at the target (see Figure 11). Charge particle density of a single ring, $\rho(R) = dQ/(2\pi R dR)$, is given by

$$\rho(R) = \rho_o \exp(-\frac{(R - R_i)^2}{2\sigma^2}), \tag{15}$$

where $R_i$ is the radius of the center of the ring, and $\sigma$ is the rms beam size. The normalization constant $\rho_o$ is determined by integration of Eq. (15) over radius:

$$Q = 2\pi \rho_o \int_{-\infty}^{\infty} \exp(-\frac{(R - R_i)^2}{2\sigma^2}) R dR, \tag{16}$$

where $Q = IT$ is the charge delivered by the beam at the target per one ring, $I$ is the beam current, $T = 1/f$ is the time required for the beam to perform one turn at the target. Within beam pulse $\tau$ the beam performs $N = f\tau$ turns at the target. Integration of Eq. (16) provides the value of the normalization constant:

$$\rho_o = \frac{I\tau}{2\sqrt{2}\pi^{3/2}\sigma R_i N}. \tag{17}$$

Charge density at the target is a superposition of overlapped circle beam distributions after $N$ turns:

$$\rho(R) = (\frac{I\tau}{2\sqrt{2}\pi^{3/2}\sigma}) \frac{1}{N} \sum_{i=1}^{N} \frac{1}{R_i} \exp(-\frac{(R-R_i)^2}{2\sigma^2}). \tag{18}$$

Beam radius at the target after $i$ turns, $R_i$, is determined from Eq. (14) as

$$R_i = R_{min}\sqrt{1+(\frac{R_{max}^2}{R_{min}^2}-1)(\frac{i-1}{N-1})}, \qquad i = 1,..., N. \tag{19}$$

The increment of beam radius after each turn, $\Delta R_i = R_{i+1} - R_i$, drops according to Eq. (10) as:

$$\Delta R_i \approx \Delta R_1 \frac{R_{min}}{R_i}, \tag{20}$$

where $\Delta R_1 = R_2 - R_{min}$ is the increment of first beam radius at the target. Because the value of $\Delta R_1$ is the largest value of increment of beam radius, the largest non-uniformity is observed at the first turn (see Figure 12). Figure 13 illustrates non-uniformity of beam distribution at the target as a function of number of turns $N$. With sufficient amount of overlapping of single beam rings, the non-uniformity quickly decreases with increasing number of overlapped circles. The raster frequency is equal to the ratio of irradiation circles to the time of irradiation:

$$f = \frac{N}{\tau}. \tag{21}$$

Assuming the number of overlapped beam circles at the target $N = 10$, and irradiation time equal to the FPNS beam pulse $\tau = 850\,\mu s$, the rastering frequency of $f = 11.76$ kHz is sufficient to achieve irradiation non-uniformity of $\Delta I / I < 1\%$. Table 7 contains the values of beam radii for target illumination using $N = 10$ turns.

An additional requirement of the rastering system is related to prevention of activation of beamline components by backscattered neutrons from the target. The target shield wall is placed 10 m upstream of the target. The diameter of the beam pipe in the shield wall should be smaller than the diameter of standard transport line beam pipe which is 6", or 15.24 cm. On the other hand, LANSCE standards require that the aperture is dimensioned such that at least 7 rms of the deliberately deflected beam clears the beam pipes. Figure 14 illustrates FPNS beam setup using rasters in front of the shield wall. For simplicity, we assume that x-y raster magnets are placed in the same position, while realistically they are supposed to be separated by the distance of 1 m. The distance from rasters to wall is 1 m, and from wall entrance to target 10 m, with wall thickness of 2 m. Taking into account variation of the beam centroid at the target 5.5 cm < $R_{axis}$ < 6.9 cm and the distance from raster magnets to target as $z = 11$ m, the beam deflection angle is varied at the time of target irradiation within an interval $dR_{axis}/dz = (5.0......6.27) \cdot 10^{-3}$. The required strength of the raster magnet of transverse field $B$ and length $L$ to provide the needed beam deflection is determined by

$$BL = \frac{mc\beta\gamma}{e}\frac{dR_{axis}}{dz}. \tag{22}$$

Figure 15 illustrates the dependence of the of raster magnet strength as a function of time within the beam pulse.

The shield wall pipe diameter is determined as

$$D(z) = 2[(\frac{dR_{axis}}{dz})_{max}(z - z_{raster}) + 7R_{rms}(z)], \tag{23}$$

where $(dR_{axis}/dz)_{max} = 6.27 \cdot 10^{-3}$ is the maximal raster angle of the central trajectory, $z - z_{raster}$ is the distance from raster magnet, and $R_{rms}(z)$ is the rms beam size. For the entrance to the shield wall, $z - z_{raster} = 100$ cm, and $R_{rms} = 0.5$ cm, the entrance diameter of the pipe $D_{entrance} = 8.25$ cm. For the exit of the shield wall, $z - z_{raster} = 300$ cm, and $R_{rms} = 0.415$ cm, the exit diameter of the pipe $D_{exit} = 9.57$ cm. Therefore, a shield wall pipe diameter of 4 inches satisfies the 7-rms clearance requirements.

## 5. Summary

This study concludes that an upgrade of the LANSCE Accelerator Facility for a 1-MW FPNS project is feasible. Expected performance of the RF systems for the Drift Tube Linac and Coupled Cavity Linac meet FPNS requirements. Development of the existing high-energy beamline to deliver 1-MW proton beam to the FPNS target requires marginal upgrade of the existing Line A. The proposed circular beam raster system provides high uniformity of irradiation area with non-homogeneity less than 1% and guarantees shielding of the beamline from backscattered neutrons.

## Acknowledgements

The authors are indebted to Daniel Rees, John Lyles, Joseph Bradley, Michael Borden, Nathan Moody and John Smedley for help and useful discussion of results.

Table 1. Current beam parameters of LANSCE accelerator.

| Area | Beam/ Energy (MeV) | Rep Rate (Hz) | Pulse Length (μs) | Chopper | Current /bunch (mA) | Average Current (μA) | Average Power (kW) |
|---|---|---|---|---|---|---|---|
| Lujan Center | H-/ 800 | 20 | 625 | 0.8 | 10 | 100 | 80 |
| Isotope Production | H+/ 100 | 100 | 625 | 1. | 4 | 250 | 25 |
| Weapons Neutrons | H- / 800 | 100 | 625 | 2.8 x 10$^{-3}$ | 25 | 4.5 | 3.6 |
| Proton Radiography | H-/ 800 | 1 | 625 | - | 10 | <1 | <1 |
| Ultra-Cold Neutrons | H-/ 800 | 10 (every 5 sec) | 625 | 0.8 | 10 | 10 | 8 |

Table 2. FPNS Beam Parameters

| | |
|---|---|
| Average beam power | 1 MW |
| Beam energy | 800 MeV |
| Average Current | 1.25 mA |
| Repetition Rate | 78 Hz |
| Pulse length | 850 μs |
| Peak current | 18.87 mA |
| Beam FWHM size at the target | 2 mm |
| Raster | Circular |
| Non-uniformity of target irradiation | < 3% |
| Number of protons at the target | 9.95×10$^{22}$ protons per year |
| Beam delivery time | 3500 hours/year |

Table 3. Beam parameters of LANSCE accelerator with FPNS.

| Area | Beam / Energy (MeV) | Rep Rate (Hz) | Pulse Length (μs) | Chopper | Current/ bunch (mA) | Average Current (μA) | Average Power (kW) |
|---|---|---|---|---|---|---|---|
| Lujan Center | H$^-$/ 800 | 20 | 625 | 0.8 | 10 | 100 | 80 |
| Isotope Production | H$^+$/ 100 | 20 | 662 -826 | 1.0 | 18.87 | 250-312 | 25-31 |
| Weapons Neutrons | H$^-$/ 800 | 100 | 850 | 2.8x10$^{-3}$ | 25 | 5.95 | 4.76 |
| Proton Radiography | H$^-$/ 800 | 1 | 850 | - | 10 | <1 | <1 |
| Ultra-Cold Neutrons | H$^-$/ 800 | 8 (every 5 sec) | 850 | 0.8 | 10 | 10.8 | 8.6 |
| FPNS | H$^+$/ 800 | 78 | 850 | 1.0 | 18.87 | 1251 | 1001 |

Table 4. RF Parameters of Drift Tube Linac (T1=Tank 1,T2= Tank 2, T3= Tank3, T4=Tank 4).

| Parameter | Value |
|---|---|
| Operation RF duty factor, % | 12.4% |
| Duty factor limit, % | 15.0% |
| Nominal power $P_{RF}$ (T1/T2/T3/T4), MW | 0.322 / 2.60 / 2.22 / 2.68 |
| Beam loading, $P_b$ (T1/T2/T3/T4) ), MW | 0.087 / 0.678 / 0.592 / 0.515 |
| Total power $P_{tot}$ (T1/T2/T3/T4), MW | 0.409 / 3.278 / 2.812 / 3.195 |
| Peak power limit (T1/T2/T3/T4), $P_{lim}$, MW | 0.425 / 3.60 /3.60 / 3.60 |
| RF FPNS pulse length, ms | 1.1 |

Table 5. RF Parameters of Coupled Cavity Linac.

| Parameter | Value |
|---|---|
| Operation RF duty factor, % | 11.4 |
| Duty factor limit, % | 12.0 |
| Nominal klystron peak power, MW | 0.75 |
| Max beam loading, $P_b$, MW | 0.30 |
| Total power, $P_{tot}$ MW | 1.05 |
| Klystron power limit, $P_{lim}$, MW | 1.25 |
| RF FPNS pulse length, ms | 0.95 |

Table 6. Parameters of beamline quadrupoles.

| Quadrupole | z-position (m) | Length (m) | Gradient (kGs/cm) |
|---|---|---|---|
| LAQM01 | 0.78542 | 0.1725 | 1.5 |
| LAQM02 | 1.03924 | 0.1721 | -1.5 |
| LAQM03 | 32.213758 | 0.6 | 0.1238 |
| LAQM04 | 33.312518 | 0.6 | -0.113 |
| LAQM05 | 64.00088 | 0.6 | 0.0859 |
| LAQM06 | 65.09973 | 0.6 | -0.1 |
| LAQM07 | 83.02873945 | 0.6 | 0.2278 |
| LAQM08 | 84.127589 | 0.6 | -0.2287 |

Table 7. Beam centroid radii for 10 beam revolutions.

| Turn, $i$ | $R_i (cm)$ |
|---|---|
| 1 | 5.50000 |
| 2 | 5.67264 |
| 3 | 5.84019 |
| 4 | 6.00305 |
| 5 | 6.16162 |
| 6 | 6.31621 |
| 7 | 6.46710 |
| 8 | 6.61455 |
| 9 | 6.75878 |
| 10 | 6.90000 |

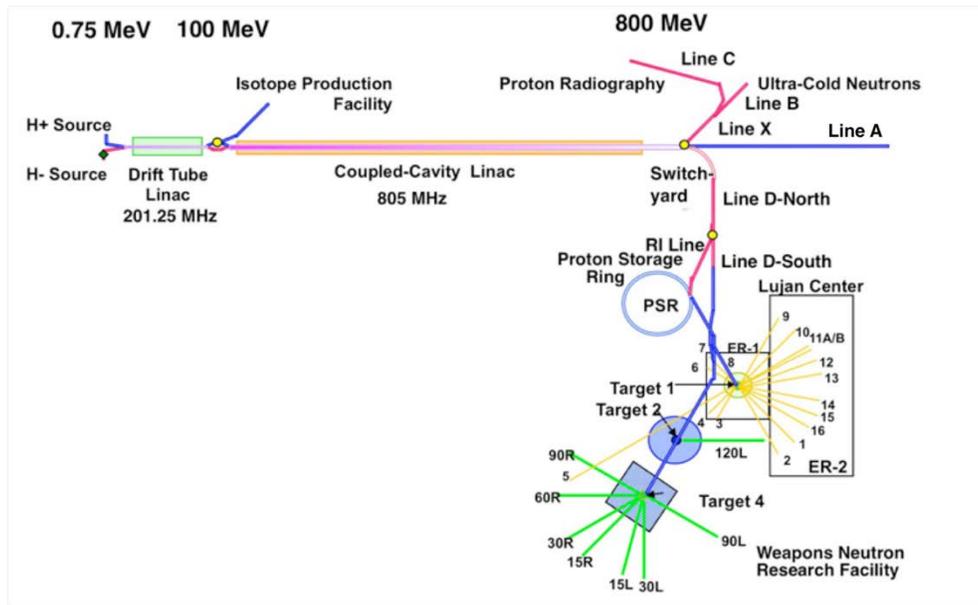

Figure 1: Layout of LANSCE Accelerator Facility.

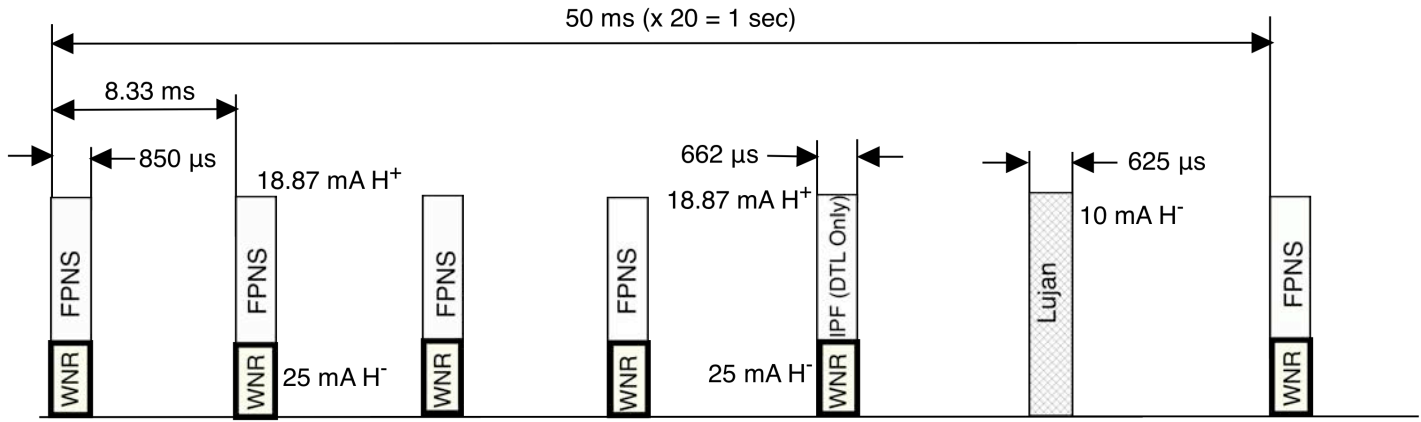

Figure 2: LANSCE beam layout with FPNS beam (WNR beam is a sequence of single 201.25 MHz bunches, separated by time interval of 1.8 µs. Parameters of other beams are given in Section 2 and Table 3).

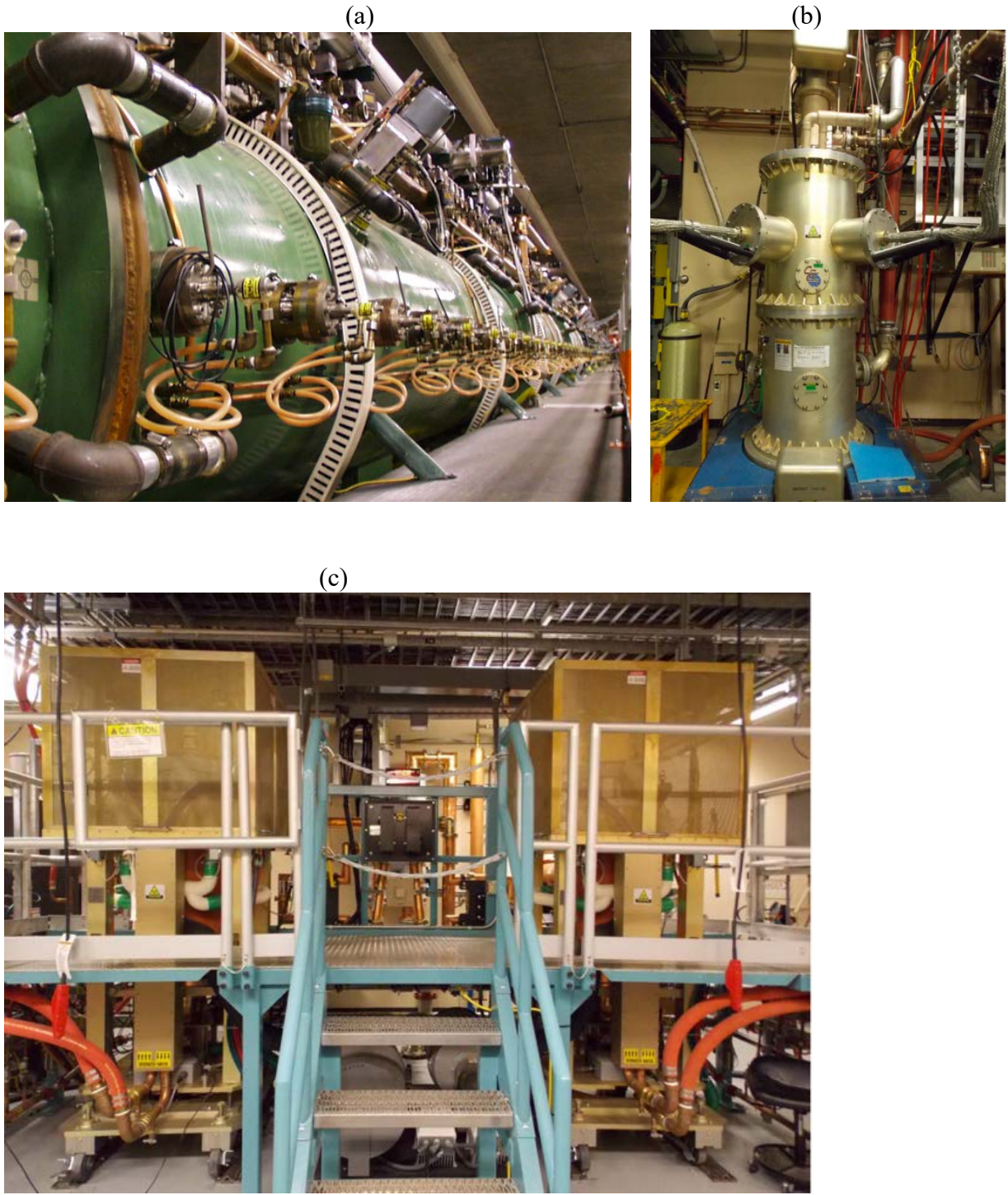

Figure 3: (a) The 201.25 MHz RF LANSCE drift tube linear accelerator, (b) Tank 1 triode amplifier, (c) Diacrodes for Tanks 2 - 4.

(a)

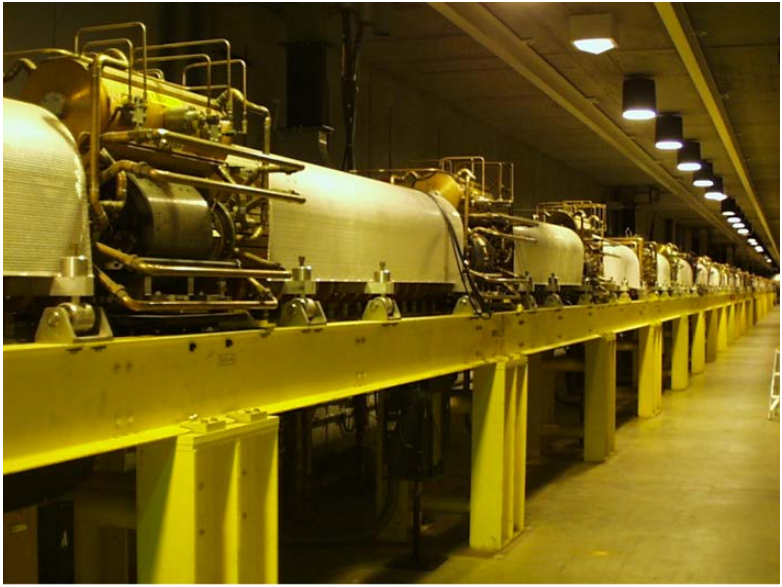

(b)

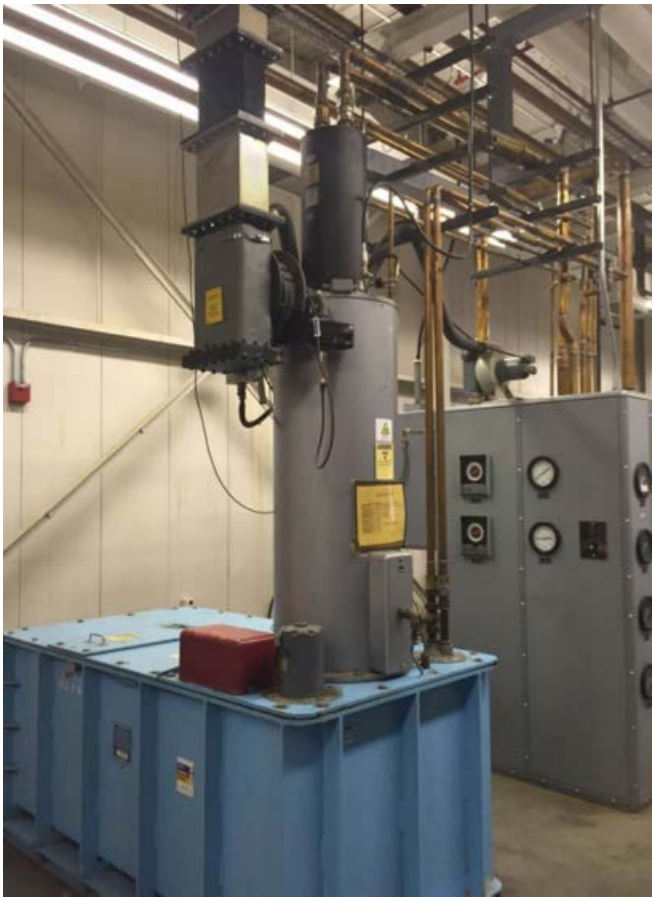

Figure 4: (a) 805 MHz LANSCE coupled cavity linear (CCL) accelerator, (b) CCL klystron.

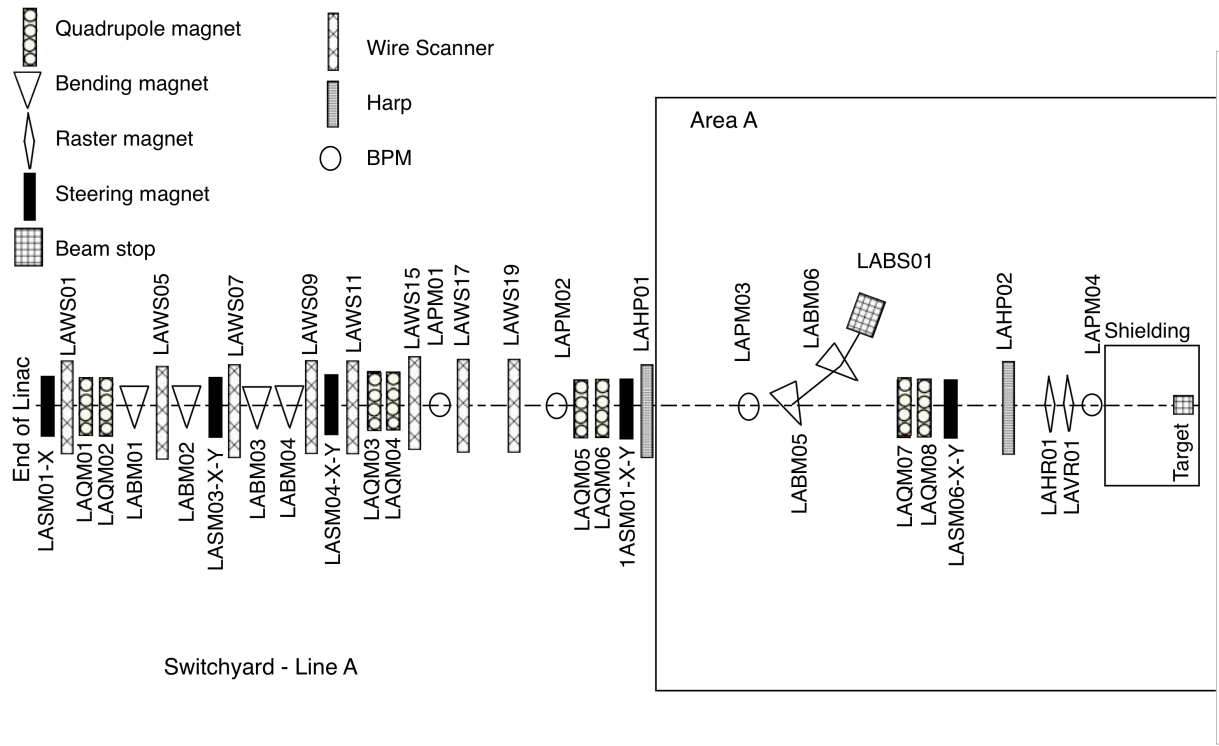

Figure 5. Beamline from the end of accelerator to the FPNS target.

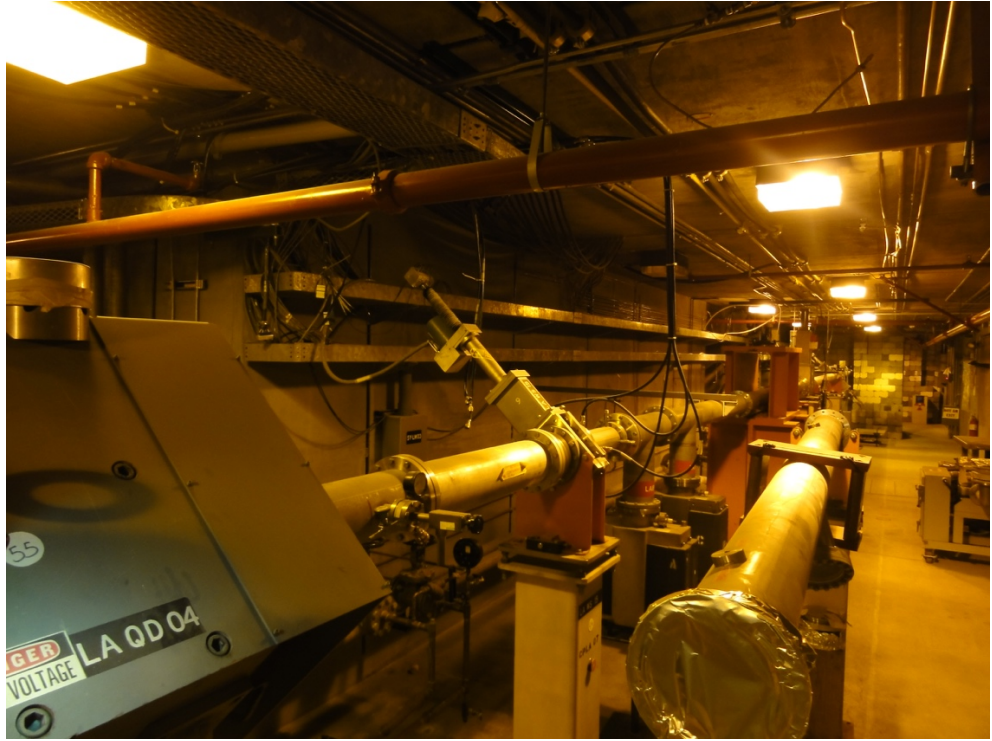

Figure 6: Line A after the linear accelerator.

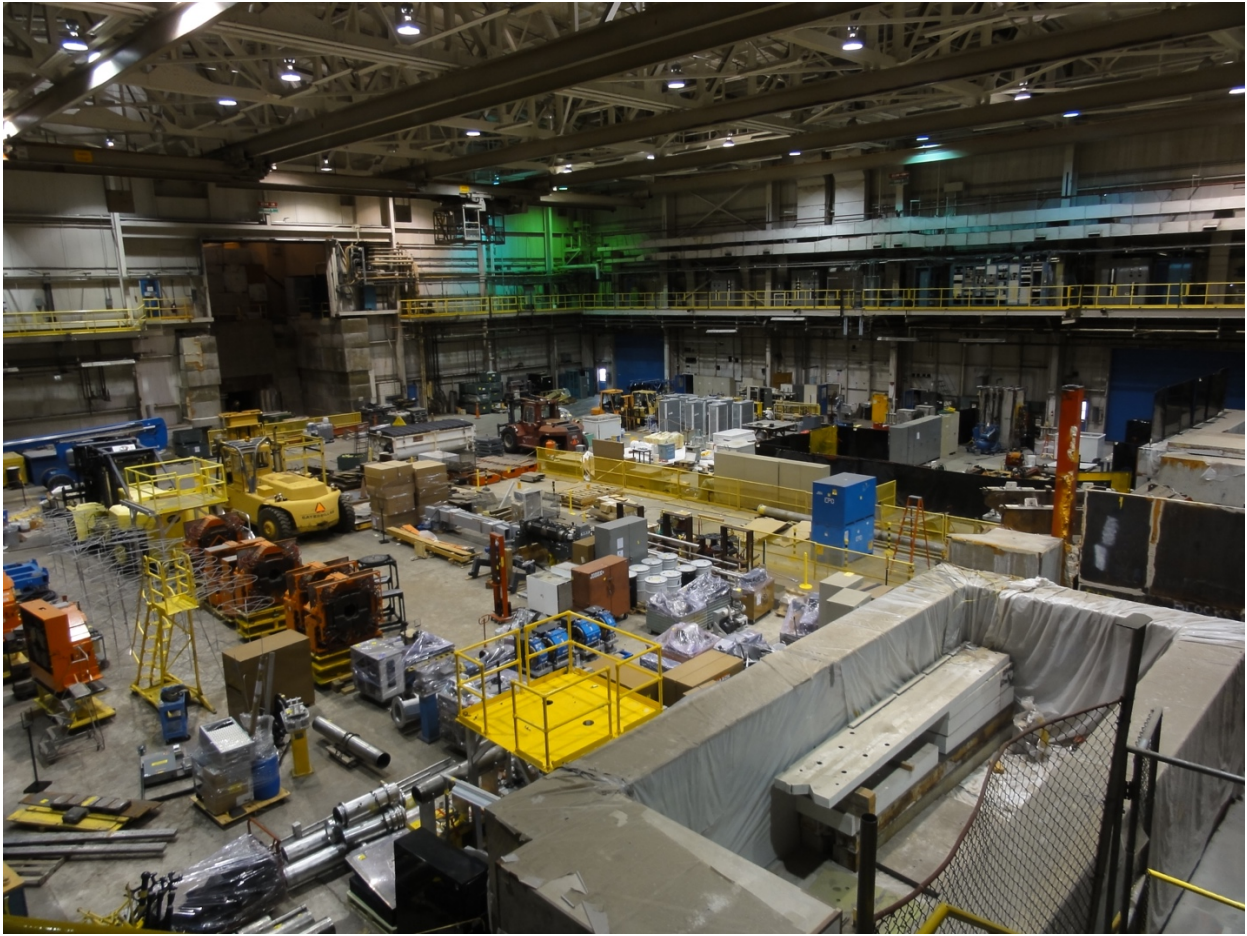

Figure 7: Experimental Area A for target installation.

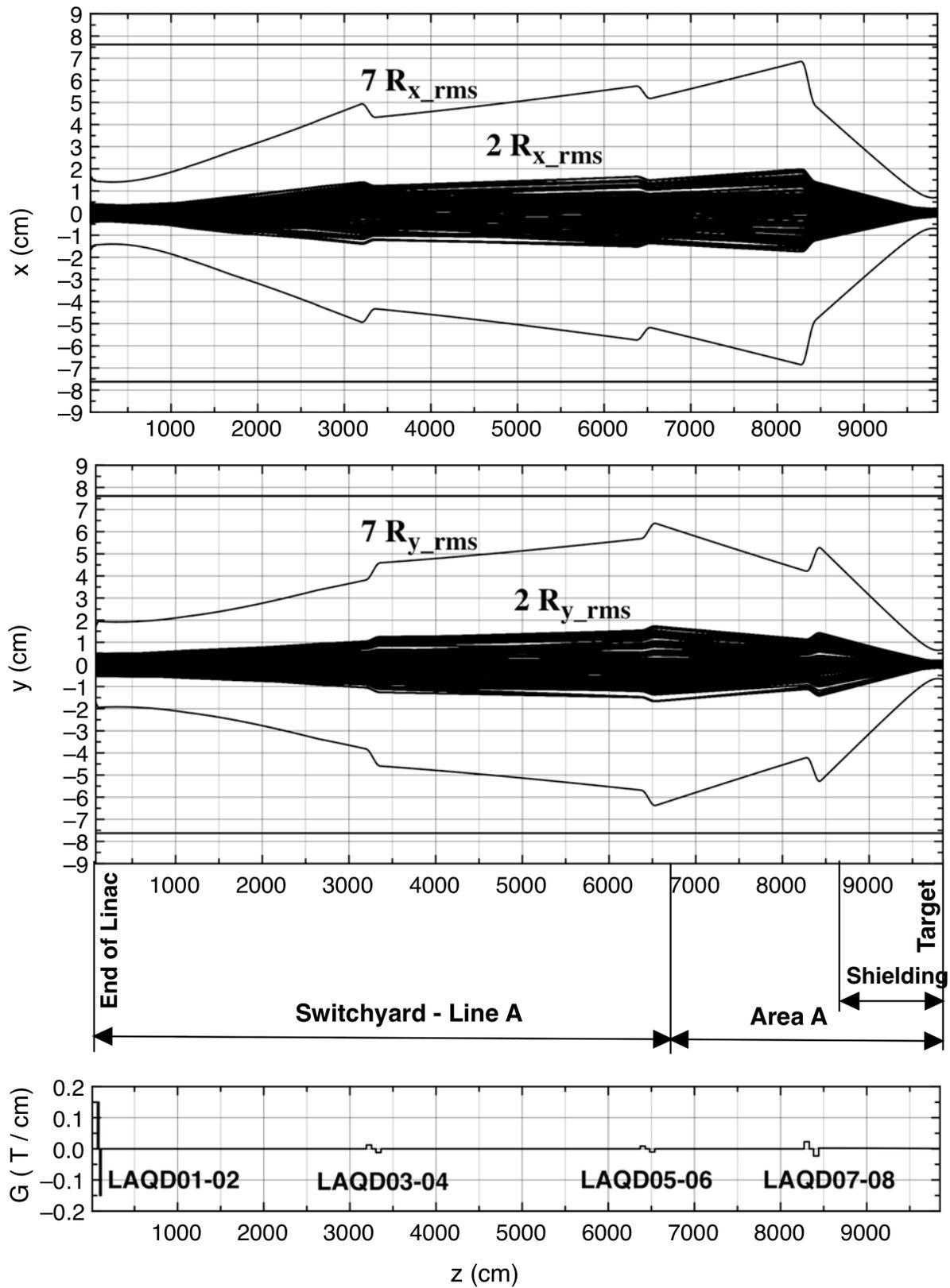

Figure 8. BEAMPATH simulation of FPNS beamline.

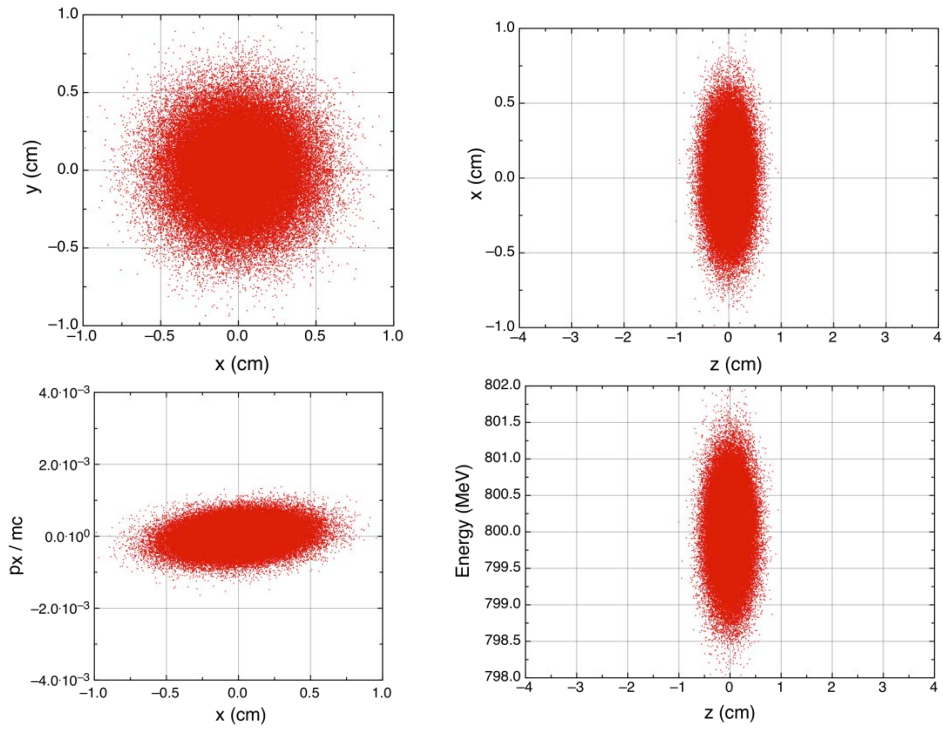

Figure 9. Beam distributions after the linac.

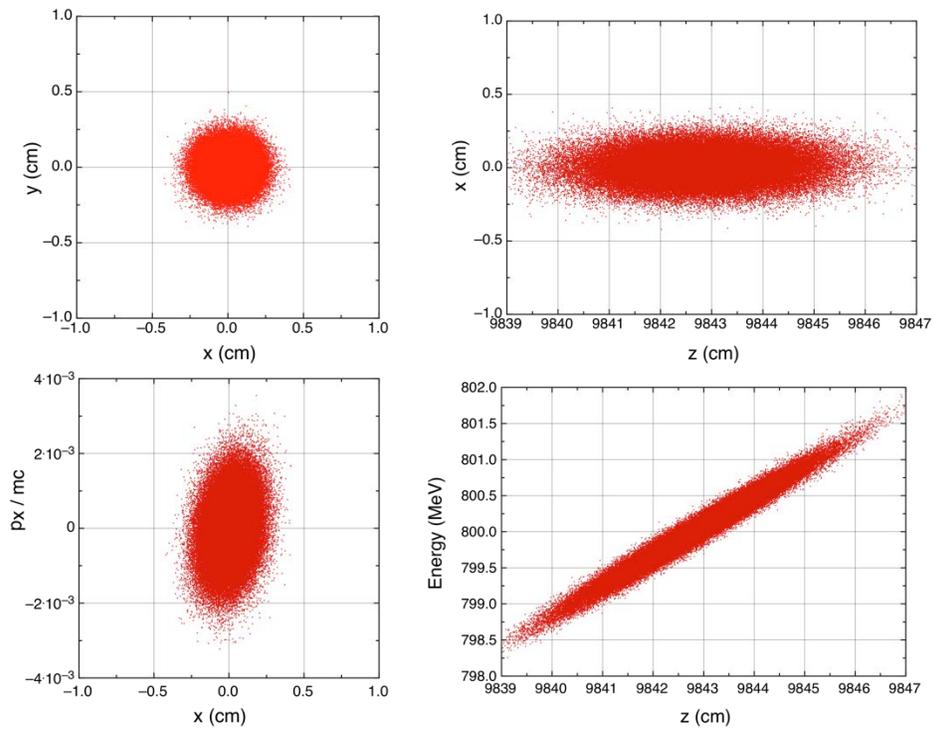

Figure 10. Beam distributions at the target.

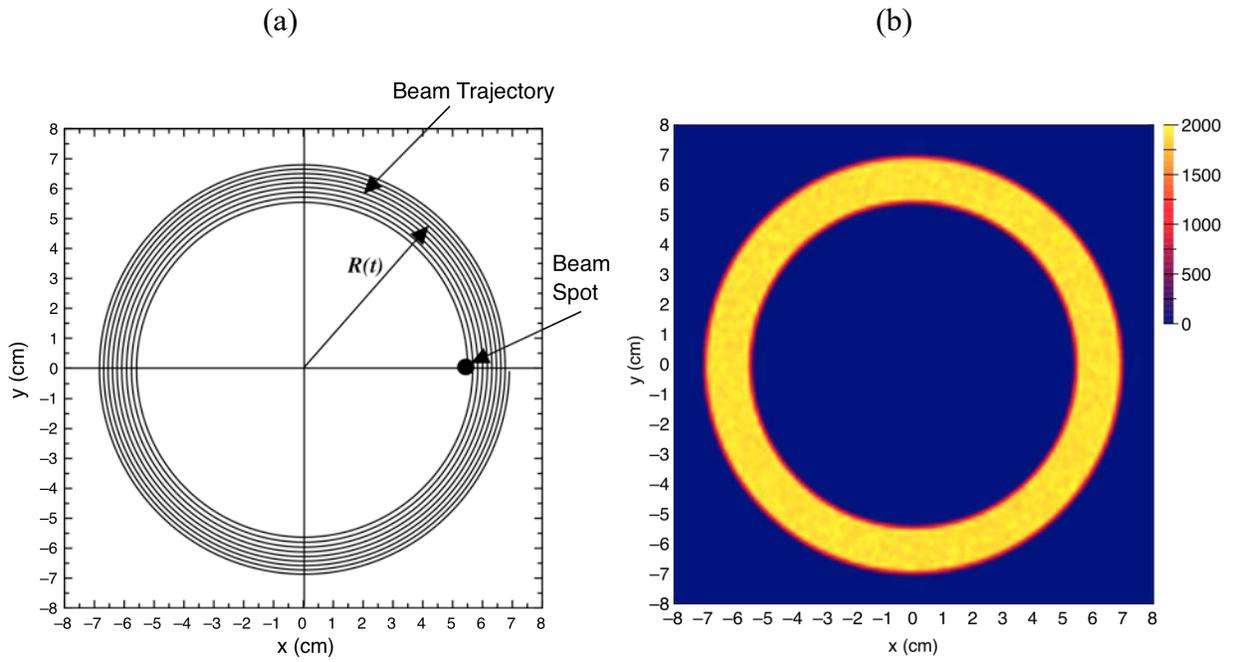

Figure 11: Irradiation of target by beam spiral rastering: (a) beam trajectory, (b) particle distribution.

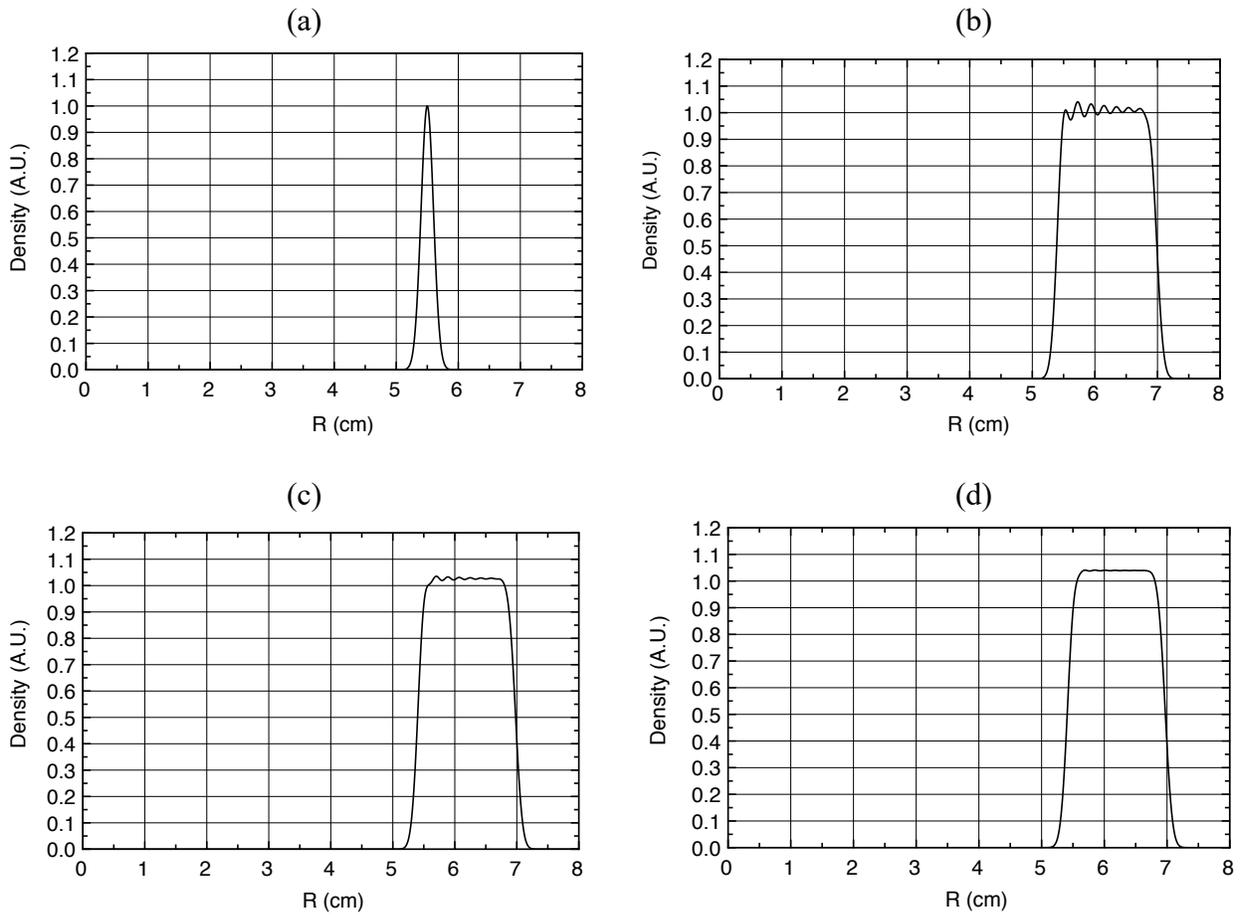

Figure 12: Distribution of particles performing circular sweeping of target within $N$ turns:
(a) $N=1$, (b) $N=8$, (c) $N=9$, (d) $N=10$.

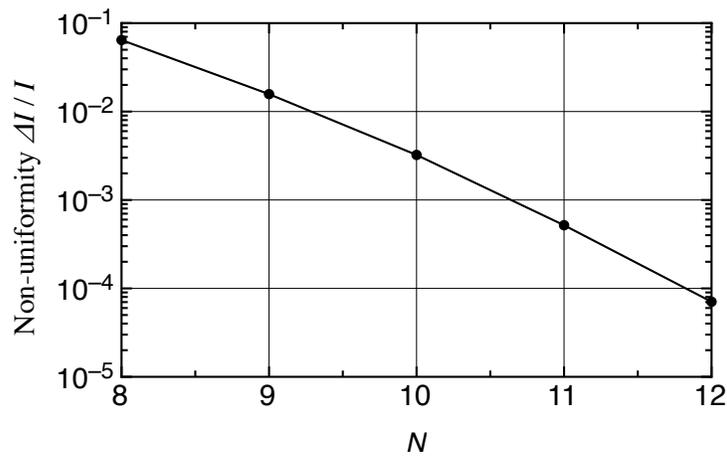

Figure 13: Maximal beam non-uniformity at the target as a function of the number of beam turns at the target.

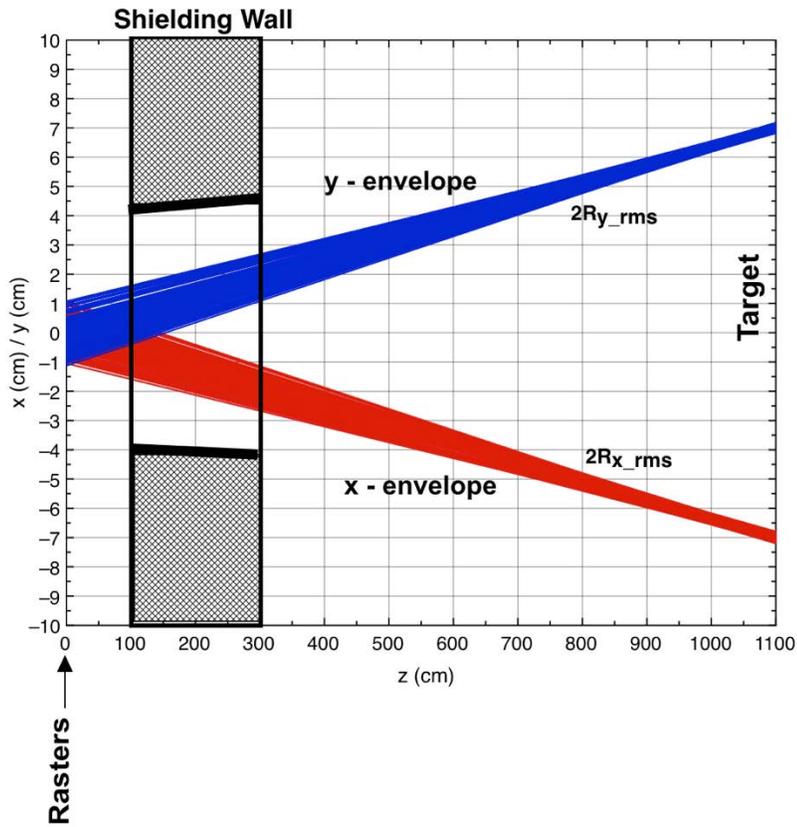

Figure 14: Beam rastering in front of the shield wall.

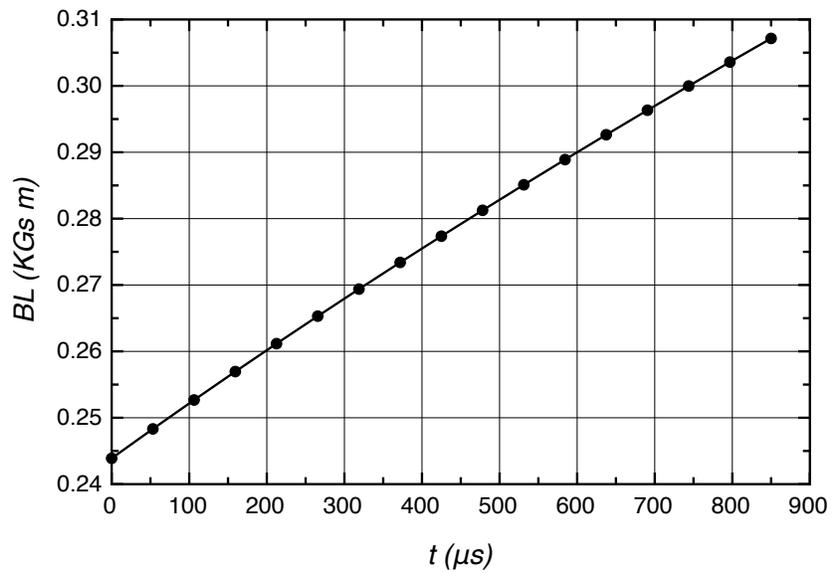

Figure 15: Variation of raster magnets field amplitude within the macro pulse.